# Securing Wireless Sensor Networks Against Denial-of-Sleep Attacks Using RSA Cryptography Algorithm and Interlock Protocol

Reza Fotohi[1]  | Somayyeh Firoozi Bari[2] | Mehdi Yusefi[3]

[1]Faculty of Computer Science and Engineering, Shahid Beheshti University, Tehran, Iran

[2]Department of Computer Engineering, Shabestar Branch, Islamic Azad University, Shabestar, Iran

[3]Department of Computer Engineering, Islamic Azad University, Dolatabad Branch, Isfahan, Iran.

**Correspondence**
Faculty of Computer Science and Engineering, Shahid Beheshti University, Tehran, Iran

Email: Fotohi.reza@gmail.com;
R_fotohi@sbu.ac.ir

**Abstract**

Wireless sensor networks (WSNs) have been vastly employed in the collection and transmission of data via wireless networks. This type of network is nowadays used in many applications for surveillance activities in various environments due to its low cost and easy communications. In these networks, the sensors use a limited power source which after its depletion, since it is non-renewable, network lifetime ends. Due to the weaknesses in sensor nodes, they are vulnerable to many threats. One notable attack threating WSN is Denial of Sleep (DoS). DoS attacks denotes the loss of energy in these sensors by keeping the nodes from going into sleep and energy-saving mode. In this paper, the Abnormal Sensor Detection Accuracy (ASDA-RSA) method is utilised to counteract DoS attacks to reducing the amount of energy consumed. The ASDA-RSA schema in this paper consists of two phases to enhancement security in the WSNs. In the first phase, a clustering approach based on energy and distance is used to select the proper cluster head and in the second phase, the RSA cryptography algorithm and interlock protocol are used here along with an authentication method, to prevent DoS attacks. Moreover, ASDA-RSA method is evaluated here via extensive simulations carried out in NS-2. The simulation results indicate that the WSN network performance metrics are improved in terms of average throughput, Packet Delivery Ratio (PDR), network lifetime, detection ratio, and average residual energy.

## 1 | INTRODUCTION

Networks constructed with multiple sensors enables novel applications covering a wide range of areas. However, it is well-known that networks forming from a range of sensors are prone to factors threatening their security [1-4].

Security is a main research topic in wireless sensor networks (WSNs). The widespread application of WSNs in security sensitive environment, such as military environment, made security considerations a basic requirement. Since nodes are the routing medium in the network, attacking the nodes eradicate the network. As routing is a trust-based operation among the nodes, there exists a good opportunity for attackers to disorder the routing process. These networks are usually formed without pre-planning and are utilised for a short period; hence, security investigations in these networks are carried out separately. Therefore, it is necessary to implement countermeasures to protect the WSN from security attacks. Denial of Sleep (DoS) attack [5-11] is one of the most notable form of attacks in network sensors. DoS refrains the radio from moving into sleep mode to completely drain its battery. In normal operating conditions, energy consumption ratio in sensors drains their battery in months, whereas denial of sleep attack drains them in a few days by keeping the radio transmitter system on the sensor nodes on [1, 12-15].

In this paper, we proposed a multistage method called ASDA-RSA, as follows: 1) dividing the network into clusters. This stage is carried out using an energy (and distance) based approach for selecting the appropriate cluster head, to improve the network and optimize energy consumption. 2) Using RSA algorithm and interlocking protocol, along with an authentication method, to prevent the DoS attack. Figure 1 demonstrate an example of DoS attack in WSNs.

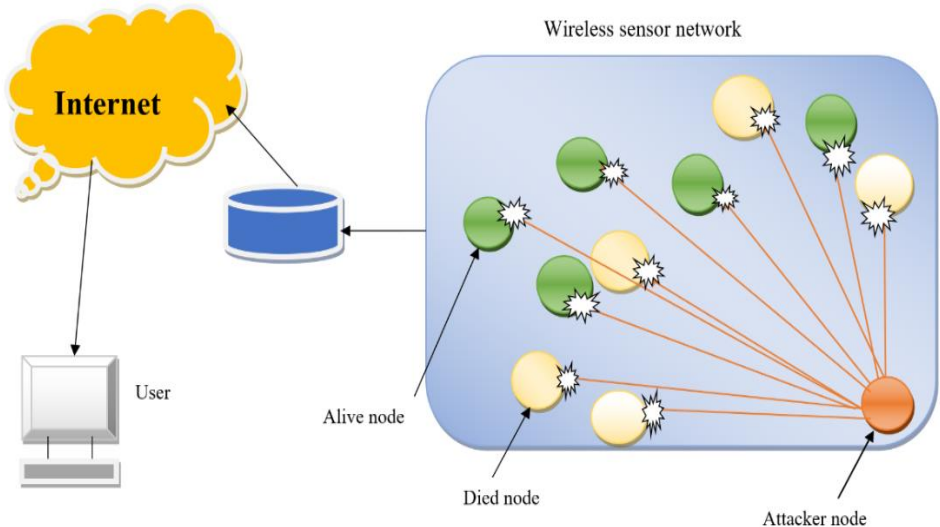

**FIGURE 1** WSNs with DoS attack [16].

The paper presented here is organized as the following. Section 2 presents the related works on approaches defending the system against Denial of Sleep attack. In section 3, cryptography algorithms for WSN are presented. In Section 4, the details of our proposed ASDA-RSA method is discussed. Moreover, parameters utilised for performance evaluation are investigated and simulation results are discussed in Section 5. Finally, in Section 6, the paper is concluded.

## 2 | RELATED WORK

Numerous studies have developed and employed a variety of security measurements to address Denial of Sleep attacks, and secure WSNs from DoS attacks. This is not a recent problem and has been previously studied extensively. There exist various techniques proposed by researchers to deal with such attacks.

In [17], the mechanisms considered in WSN MAC systems that act as countermeasures against attacks that target MAC layers in WSNs are studied. Therefore, a secure hybrid MAC mechanism entitled "Green and Secure Hybrid Medium Access Control (GSHMAC)" is proposed here to mitigate the effects such attacks have on WSN MAC systems. Threshold-based MAC mode collision control and countermeasures on WSN MAC systems with the help of internal MAC mechanisms, are instances of features provided within this mechanism.

In [18], through inspiration from the immune system of the human, an IDS framework is proposed to be applied to the context of the wireless sensor networks. A revised decentralized and customized version of the Dendritic Cell Algorithm is employed in this algorithm to provide nodes with capabilities such as neighborhood monitoring and collaboration towards identifying intruders. Moreover, to demonstrate the efficiency in terms of denial-of-sleep attack detection as well as energy consumption, the results are compared with a Negative Selection Theory implementation.

Two major contributions are offered in [19]. First, the existing defense mechanisms for the denialof-sleep attacks in ContikiMAC are tailored to CSL, due to the superiority of CSL over ContikiMAC. Second, several enhancements in security are proposed for the state-of-the-art defense mechanisms against denial-of-sleep. As demonstrated, the defensive mechanisms we proposed for CSL against denial-of-sleep reduce the number of denial-of-sleep attacks drastically, while enabling protection against a larger range of denial-of-sleep attacks compared to the existing defensive mechanism in ContikiMAC.

The vulnerabilities existing in MAC solutions as well as embeddable mechanisms to provide inherent security are the main focus of this paper [20]. Specifically, we investigate scenarios of physical jamming in which, attackers prevent communication establishment and thus, result in repetitive retransmissions as well as additional duty for the target devices. Prior to focusing on technologies in communication that are investigated by a number of main standardization bodies, numerous existing attacks are investigated in detail. Moreover, time-synchronized and channel hopping (TSCH) networks, which are designed for industrial wireless devices, are considered. The solutions provided result in collision risks minimization and idle listening reduction, while providing cryptographic suites to guarantee additional authentication and encryption.

To detect DoS attack in WSNs, a mathematical model is developed in [21] based on absorbing Markov chain, where the sensor nodes were detected using the Markov chain model. Using this model, attacks are identified by considering the expected time of death in sensor networks with a common scenario.

In [22], a distributed cooperation-based hierarchical framework is employed to detect denial of sleep attack in WSNs. This technique provides an efficient heterogeneous wireless sensor network with reliable performance by detecting anomalies in two stages, to minimise the likelihood of improper intrusions. To reduce the risk of attack, this study isolates the networks from malicious nodes to deny receiving fake packets.

Compact hierarchical model stated in [23] is used in heterogeneous wireless sensor networks to detect sleepless nodes affected by the attack. In this method, a cluster based on effective energy is used to create a five-layer hierarchical network to increase its scalability and longevity. In this approach, energy efficiency has been achieved by minimising the number of active sensors. Moreover, the designed dynamic model protects the sensor nodes from sudden death. This model uses an intrusion detection system through power analysis responsible for detecting tasks. Anomaly detection method is employed in approaches to avoid intrusion detection.

The swarm based defensive technique proposed in [24] copes with denial of sleep attacks by employing anomaly detection model to determine the traffic impact among sensor nodes. Ant factors are used as swarm information to collect oscillation and communication frequency. The faulty channel is detected according to the frequency of the oscillation; and when the node is isolated, data is achieved, and the faulty channel is removed.

In [25], intrusion detection system detects the malicious node in an isolation table based on the attack behaviour. The sensor node behaviour is compared against the attack behaviour to specify abnormal information. If the node performance became abnormal, it is separated and registered in the isolation table. In Isolation Table Intrusion Detection System (ITIDS), sensor nodes are responsible for monitoring each other and detecting denial of sleep attack.

Storm control mechanism defined in [26] is utilized to reduce data flooding and avoid denial of sleep attacks. The system tracks the frequency of the packets received, and when the expected frequency configuration is exceeded, the node alerts the base station and shuts down its wireless receiver for a pre-defined period of time.

A two-tier system based on a secure transmission scheme (CrossLayer) is proposed in [27]. In the proposed scheme, a hash-chain is employed for generating dynamic session keys for symmetric encryption key and performing mutual authentication. This scheme requires only some simple and fast computations in dynamic session key for hash functions, such as MD5 or SHA-1. Since this scheme is integrated with MAC protocol, no extra packets exists in comparison with the current MAC designs. Moreover, detailed energy distribution analysis reveals a potentially novel decision rule to compromise the energy conservation and security scheme trade-off.

In [28], proposed two optimizations to Contiki MAC. The dozing optimization, on the one hand, significantly reduces the energy consumption under ding-dong ditching. Beyond that, the dozing optimization helps during normal operation as it reduces the energy consumption of true wake ups, too. The secure phase-lock optimization, on the other hand, is a version of Contiki MAC's phase-lock optimization that resists pulse-delay attacks. Additionally, the secure phase-lock optimization makes ContikiMAC resilient to collision attacks, as well as more energy efficient.

In [29], Gunasekaran et al. suggested a denial of sleep attack detection (GA-DoSLD) algorithm based on an efficient Genetic Algorithm (GA) scheme to analyse the nodes unnormal behaviour in the nodes. In their suggested algorithm, a Modified-RSA (MRSA) algorithm is implemented in the base station (BS) to generate and distribute key pairs amongst the sensor nodes. Prior to transmitting the packets, an optimal route using AODV protocol is determined in the sensor nodes. Following determining the optimal route, the trustworthiness of the relay node is ensured using fitness calculation.

Keerthana and Padmavathi [30] suggested an Enhanced Particle Swarm Optimization (EPSO) technique for detecting the sinkhole attacks in WSN. When compared to the existing ACO and PSO algorithms, the suggested algorithm provided optimal packet delivery ratio, message drop, average delay, and false alarm rate.

In [31], a two-phase system is proposed to detect wormhole attacks and protect in the MANET, DAWA. The proposed methodology is comprised of two phases. Initially, the efficient routes are selected in the system via fuzzy logic; then, artificial immune system is utilized to identify the immune route in the selected routes. In this paper, constructing the wormhole tunnels using out-of-band high power channels results in efficient wormhole attacks identification. However, there is a prerequisite consideration in this paper that when the destination nodes are affected by wormhole attacks, they do not receive any packets, which is necessarily not true.

A transmission/reception device with wake-up radio for a node with limited resources such as an IoT network node. The device includes a permanently powered auxiliary circuit, capable of detecting a wake-up token, and the main circuit, normally in the idle state and activated by the auxiliary circuit when a wake-up token is detected. The next wake-up token is calculated by the main circuit by applying a one-way function to at least part of a message exchanged on the main radio through a secure communication [59].

In [60], the authors investigated existing studies and provided a systematic review of EDAs and defenses in LPW networks. This paper also indicated the security challenges in LPW networks related to EDAs along with the potential research directions. While LPW technologies have already hit the market with the promising deployment schedules, our attempt can inspire the research community to enhance the security of underlying protocols that will shape the connectivity of billions of devices in the future IoT ecosystem.

In [61], authors addressed the problem of secure data transmission and balanced energy consumption in an unattended wireless sensor network (UWSN) comprising of multiple static source nodes and a mobile sink in the presence of adversaries. The proposed system comprises of three phases: the identification of data collection points (convex nodes), path planning by the mobile sink, and secure data transmission. An energy-aware convex hull algorithm is used for the identification of data collection points for data transmission to the mobile sink. Data are securely transmitted through elliptic curve cryptography based ElGamal scheme for message authentication. A data packet is associated with a digital signature.

Wireless sensor networks which form part of the core for the Internet of Things consist of resource constrained sensors that are usually powered by batteries. Therefore, careful energy awareness is essential when working with these devices. However, the absence of security protection could give room for energy-drain attacks such as denial-of-sleep attacks which has a higher negative impact on the life span (availability) of the sensors than the presence of security techniques. This paper focuses on denial-of-sleep attacks by simulating three Media Access Control (MAC) protocols - Sensor-MAC, Timeout-MAC, and TunableMAC - under different network sizes. We evaluate, compare, and analyse the received signal strength and the link quality indicators for each of these protocols [62].

## 3 | Cryptography Algorithms for WSN

Many schemes based on public-key cryptography are investigated. Those algorithms are DES[1], 3-DES[2], AES[3], Blowfish, and RSA[4]. This paper consists of RSA algorithm, which can be used for key distribution and decrypting the message.

### 3.1 | DES Algorithm

The DES algorithm encrypts data with a length of 64 using a 56bit key and uses the remaining 8 bits for checking the parity. This algorithm is a symmetric block encryption algorithm and uses the Feistel cipher property. This algorithm includes 16 round and each round is called a cycle. This means that its main algorithm gets repeated 16 times in order to generate the encrypted data. The advantage this algorithm has is that due to its low-key length, encryption and decryption is done with higher speed. However, because of the low length of the key and its easier decryption by perpetrators, it was not welcomed and the AES algorithm replaced it as the security standard.

---

[1] Data Encryption Standard
[2] Triple DES
[3] Advanced Encryption Standard
[4] Rivest Shamir Adelman

### 3.2 | 3-DES Algorithm

The 3DES algorithm, which is symmetric, uses three keys and three repetitions of the DES algorithm. The function follows an encryption-decryption-encryption (EDE) sequence. 3DES has a key with a 168bits length. This algorithm has been carefully studied during its time more than any other algorithm and no vulnerability other than the meet-in-the-middle attack has been discovered. Therefore, there is high confidence in the resistance of the 3DES algorithm against password discovery. The main problem with this algorithm is that first, it is weak in terms of software and second, it has three times as many executive operations as the DES algorithm and therefore, it will definitely be slower. The second problem is that 3DES uses a data block with a 64bit length. Because of these problems, it was seen in the long term that 3DES is not a logical candidate. Finally, AES replaces 3DES which has a significant efficiency and security capability.

### 3.3 | AES Algorithm

This algorithm is a symmetric encryption algorithm which uses the same key for encryption and decryption. AES encrypts data in 128bit block and uses 128bit, 192bit, or 256bit keys. The main attack in the case of AES is related to the management of the encryption key. In case of weakness in the management and maintenance of this key, it will be possible to steal it and access the information. This algorithm is a method for the encryption of digital data and has replaced the digital encryption standard (DES). The AES standard is based on a design principle called substitution permutation network and is quick both on software and hardware. AES is a block cipher algorithm but unlike the DES standard, it does not use the Feistel cipher. In each round, AES uses the data block in parallel and performs the substitution and relocation on it. The size of the key used in AES determines the number of conversion cycles which converts the input named plaintext to the output named ciphertext.

### 3.4 | Blowfish Algorithm

This algorithm is an encryption method which has been based on a cipher key and a fast and symmetric block cipher. It is easily implemented on fast 32bit processors and needs less than 5k of memory. Blowfish presents a good encryption rate in software and to this date, no effective cryptanalysis has been discovered on it. The length of the block used in this algorithm is 64bits while the length of the key is variable in this algorithm and can be from 3 to 448 bits. This algorithm consists of two sections: key expansion and data encryption. Key expansion transforms a key of variable length (56 bytes maximum) to an array of several subkeys with an overall size of 4168 bytes. The encryption step has 16 rounds. Each round consists of a permutation, based on the key and another permutation, based on the key and data. This algorithm, by accepting a public key with a length of

32 to 448 is much faster than other methods like DES and is a good alternative. Also, by having a variable key length and increasing it, brute force attack is impossible on this algorithm.

### 3.5 | RSA Algorithm

Employing encryption based on privacy homomorphic provides simultaneous end-to-end confidentiality and data aggregation. Two instances of privacy homomorphism are called additive PH and multiplicative PH [32]. An additive homomorphic encryption algorithm supports additive operations on encrypted data without the need to decrypt any individual data. i.e. $ie.E(x+y) = E(x)+E(y)$. Operation expenses for additive PH is less than multiplicative PH, hence it is more suitable to utilise in wireless sensor networks [32]. On the other hand, a multiplicative homomorphic supports multiplicative operations over encrypted data without the need to decrypt individual data. $ie.E(x*y) = E(x)*E(y)$. Privacy homomorphic supporting cryptographic algorithms are divided into two categories: symmetric PHs [33], Asymmetric PH/public key encryption. RSA cryptography algorithms is described as follows:

#### 3.5.1 | Key generation steps of RSA

Algorithm 1 presents the pseudo code for generating RSA.

---
**Algorithm 1** Pseudo code for RSA Key generation
---
1: **Initialization:**
2: **Procedure** Key generation
3: **Step 1**: Choice 2 large prime numbers $S$ and $R$, where $S \neq R$.
4: **Step 2**: Compute the value of $m$.
5: $\qquad m = S*R$, $n$ will be used as the modulus for both public and private keys
6: **Step 3:** Find the totient of $m, \delta(m)$
7: **Step 4:** Evaluate $\quad \delta(m) = (S-1)*(R-1)$.
8: **Step 5:** Choose an integer e in such a way that $1 < i < \delta(m)$, and in such a way that $i$ and $\delta(m)$ no divisors other than
9: $\qquad gcd(i, \delta(m)) = 1$.
10: **Step 6:** Compute the value of $di$ based on relation,
11: $\qquad di \equiv 1+k\ \delta(m)$
12: $\qquad$ keep $di$ as a private key.
13: **End Procedure**

---

Public Key is $(\delta, m)$: public key is available to both cluster members and CH.

Private Key is $(di, m)$: private key is available to sink or base only.

#### 3.5.2 | Encryption

Algorithm 2 shows the pseudo code for Encryption Key.

**Algorithm 2** Pseudo code for RSA Encryption Key

1: **Initialization:**
2: **Procedure** Encryption
3: **Step1**: Sensor nodes have public key $(i, m)$ [$i$ is public]
4: **Step2**: $Cf = (M \wedge i) \mod (m)$.
6: **Step3:** The packet is encrypted
13: **End Procedure**

### 3.5.3 | Decryption

Algorithm 3 shows the pseudo code for Decryption Key.

**Algorithm 3** Pseudo code for RSA Decryption Key

1: **Initialization:**
2: **Procedure** Decryption
3: **Step1**: Base Station (BS) has private key $(di, m)$ [$di$ is private key]
4: **Step2**: $M = (Cf \wedge di) \mod (m)$.
6: **Step3:** The packet is decrypted
13: **End Procedure**

## 4 | Abnormal Sensor Detection Accuracy (ASDA-RSA)

In the following section, we design a Denial of Sleep-immune schema by employing the RSA algorithm and the interlock protocol.

### 4.1 | System architecture

The wireless sensor network in the proposed ASDA-RSA consists of nodes distributed randomly. Fig. 2 illustrates the base station and node positions our proposed ASDA-RSA architecture. We consider the following specifications for our proposed network:

- All nodes are fixed and the base station receives information from the head cluster.
- The BS is connected to the Internet.
- All nodes are homogeneous and are bound or restricted in energy.
- The nodes are not aware of their position. But each node can calculate its own distance with another node according to the received message signal strength.
- The nodes continuously sense the surrounding environment and send the relevant data at a constant rate.

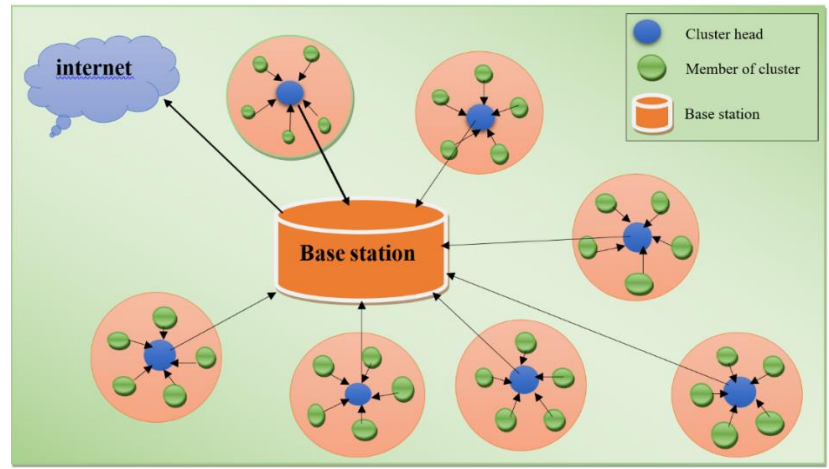

**FIGURE 2** Base station and nodes position

## 4.2 The Model for Energy Consumption

Major energy consumption in WSNs is to perform sensing, processing the data, and communication. In this paper, first order ratio model is utilised as the energy consumption model, and energy consumption is considered for data communication only. Relative to the sender and receiver nodes distance ($di$), energy consumption will be modelled as a free space or a multipath fading channel. In Eq. (6) [34], the amount of energy consumed to transmit $T$ bits of data to distance $di$ is demonstrated:

$$E_{tx}(T, di) = \begin{cases} T * E_{elec} + T\varepsilon_{fs} * d^2 & di < di_0 \\ T * E_{elec} + K\varepsilon_{mp} * d^4 & di \geq di_0 \end{cases} \quad (6)$$

In this equation, $E_{elec}$ is the amount of energy required to activate the circuits, $\varepsilon_{mpi}$ and $\varepsilon_{fsi}$ are relay activation energy model for multi-path channels and empty space, respectively, $di^2$ denotes the power loss in free space, while power loss in multipath fading is $di^4$. Currently, there is a great deal of research in the area of low energy radios. In our work, we assume a simple model where the radio dissipates $E_{elec} = 100 nJ/bit$ in the transmit or receive circuitry and $\varepsilon_{mpi} = 0.0015 pJ/bit/m^4$ for the transmit amplifier.

Simulated parameter is set as: $\begin{cases} E_{elec} = 100 nJ/bit, \\ \varepsilon_{fsi} = 20 pJ/bit/m^2, \\ \varepsilon_{mpi} = 0.0015 pJ/bit/m^4 \end{cases}$

Furthermore, $di_0$ Eq. (6), which is a constant value, is calculated according to Eq. (7) [34]:

$$di_0 = \sqrt{\frac{\varepsilon_{fs}}{\varepsilon_{mp}}} \quad (7)$$

Plus, the necessary energy to receive $T$ bits of data is as demonstrated in Eq. (8) [34]:

$$E_{Rx}(T, di) = T * E_{elec} \quad (8)$$

It is assumed that every cluster head receives a single packet from its cluster members in each communication period. Once all packets are received, the useful information are combined into a single packet in the cluster head and are reported to the sink via a multi-hopping communication. The amount of energy consumed to combine data at each cluster head node is calculated according to Eq. (9):

$$Eda_v = NM * k * EDA \quad (9)$$

Where $NM_v$ is the members' count in the cluster head $v$, and $EDA$ is the combination cost per data bit [34].

## 4.3 Cluster Head Election in the Proposed ASDA-RSA

Every node in wireless sensor networks have certain energy levels and possess certain geographic coordinates. Regarding the characteristics of such networks and the significance of their energy, if a network node can be selected as a head cluster to decrease the required space compared to the sink, and consumes more energy, then the network would have a longer life span. To aggregate the data on a given network, it is better to analyse the data individually in each node, and transmit the results to the cluster head in order to reduce computation burden for each cluster head. In sensor networks, if the node count is less than five, clustering is not mandatory, and data are transmitted directly to the sink. In our proposed ASDA-RSA method, appropriate cluster head to collect data is selected initially. The steps to select the cluster head is as the following:

**Step 1:** First, it is best to select a node closer to the base station. Euclidean Eq. (1) is employed for this. The nodes in the network are all assumed to be equipped with a Global Positioning System (GPS), and is aware of its geographic location.

$$(I.S) = \sqrt{(x_i - x_s)^2 + (x_i - x_s)^2} \tag{1}$$

Where $I$ is the coordinates of each node, while $S$ is the sink coordinate.

**Step 2:** Next, the nodes that are more energy-efficient among the selected nodes, and are closest to the base station are more suitable for the cluster heads. According to Eq. (2), the energy of the nodes are measured as the following:

$$Q = \frac{e}{t} \tag{2}$$

Where $e$ represents the node initial energy in kilojoules, while $t$ is the time parameter in minutes. Eq. (2) can be rewritten by converting the values in terms of Jules and seconds (in the first round all nodes have homogeneous and identical energies).

**Step 3:** Following calculating the energy for the nodes, and determining the nodes and the base station distance, Eq. (3) is employed to select the cluster head.

$$r_i = e_i * d(I.S) \tag{3}$$

In Eq. (3), $e_i$ represents the energy of the nodes, while $d(I.S)$ is nodes and the sink distance. Each node transmits its calculated energy value to the neighbouring nodes closest to the sink. Each node compares the $r_i$ of its neighbouring node with its own, and if their $r_i$ is less than their neighbours, the node with the lesser energy is a normal node.

**Step 4:** To select the cluster head, the energy consumed in each node should also be calculated in the fourth step, so that we could opt the node with the maximum available energy as a cluster head. For this purpose, we use Eq. (4):

$$e = Q * t \tag{4}$$

Finally, the nodes with the most remaining energy and the least distance to the sink are selected to be cluster head nodes. The cluster head selection is such that all nodes send off their remaining energy and distance to the nodes near the sink, nodes with the most energy introduce themselves as the cluster heads, informing their cluster members by sending a message. In the next section, a process is discussed to prevent Denial of Sleep attacks.

## 4.4 | Preventing Denial of Sleep attacks in ASDA-RSA

Our proposed method is implemented on S-MAC protocol to demonstrate detection and preventing Denial of Sleep attacks. To prevent such attacks, we use RSA algorithm and interlock protocol to perform key exchange, plus a method for node authentication. The procedure to transfer the keys to other nodes is a vital process, since keys are prone to attacks. To protect the keys against such attacks, a key transfer operation is performed using an interlocking protocol, where AES algorithm is employed to perform key encryption. The size of the key utilised in the AES code determines the number of repetition cycles in the conversion, which converts the input (plain text) to the output (encoded text). Each iteration has several processing steps, depending on the encryption key. On the receiving nodes, a set of reverse cycles convert the encoded text to the original text using a similar encryption key.

In the interlocking protocol, the locked key splits into two parts. The transmitter sends the first part of the key at the beginning of the process. The second part however, is transmitted when a response is received from the receiving node. It is apparent that the key can be decrypted at the recipient base station only if the two key parts are both received and joined together. To do this, the nodes connecting to the network must agree with the key symmetric encryption technique. In our proposed method, we employ AES algorithm, which divides the encryption key into two parts. For the transmitting protocol, the second part is transmitted, following the authentication that it is transferred to the correct nodes. In sensor networks, S-MAC protocols set the sleep periods. In these protocols, synchronisation signals are transmitted to adjust the node clocks. In these protocols, a large number of control messages, such as RTS and CTS, are employed, which are referred to as synchronisation packages ($SYN$). An effective technique to perform Denial of Sleep attacks is repeating control packets such as RTS messages, to prevent the nodes from going to sleep mode, resulting in waste of energy. A transmission stream of such messages with short intervals will not give the network nodes enough idle state time to initiate sleep mode, thus leading to battery power loss. While the attack is ongoing, all nodes can be affected by this damage. The attacker transmits $SYN$ messages, indicating the next sleep interval for the nodes. When a $SYN$ packet is received with similar sleep timing from another node, the upcoming sleep time is calculated to maintain the synchronization. In sensor networks, sleep times are not simply resettled, instead their times are received from the $SYN$ packets. Eq. (5) calculates the $New_{SleepTime}$ value.

$$New_{SleepTime} = \left( Old_{SleepTime} + \frac{\text{Re}ceived\ SYN\ Packet\ Sleep\ Time}{2} \right) \tag{5}$$

According to this method, sleep timing is not affected abruptly with the receipt of a $SYN$ message. In fact, this method enables gradual synchronisation for the nodes, with a similar program. A Denial of Sleep attack can also perform via repeating the $SYN$ packets. The main reason is that the attacker, who is monitoring the network, can easily identify these packages; even if they are encrypted. To do this, the attacker should merely monitor all messages for their size and time. For instance, the size of S-MAC $SYN$ packets are ten bytes. Plus, the framing for these packets in these S-MAC frames are mainly placed in the first milliseconds. If the attacker obtains this information, the package manipulation is accomplished regardless of its encryption. Therefore, in this paper we employ authentication to protect the network. In algorithm (4), the Pseudo code for our proposed ASDA-RSA schema is demonstrated.

*Cost analysis of the algorithm 4*: In the proposed method, because of using the RSA cryptography algorithm, the communications between the sensors and also the packet transmissions are carried out with higher security. However, due to the high key length, encryption and decryption costs are high. Also, the authentication phase is carried out in two sensor authentication and network authentication phases which has a higher computational cost. But the overall computational cost of the proposed method is less than other methods. However, the memory usage of the proposed method is since it contains the packet information for authentication in the cluster and network levels along with the cluster information and their distance from the central sink.

---

**Algorithm 4** Pseudo code for ASDA-RSA proposed schema

```
1:  Step 1. Network Clustering
2:         Create the random WSN as sensor nodes;
3:         Define parameters of Algorithm 1 by Table 2 and Table 4;
4:         Select the cluster head (CH);
5:         Calculate the energy of each node;
6:         Calculate the distance of each node;
7:         Selecting of a node with a low distance to the base station and high energy as a CH;
8:  Procedure Node Authentication
9:  Step 2. Node Authentication at cluster level
10:        Calculate Threshold;
11:        Calculate Interval;
12:        While Interval > Threshold  do
13:                For All of the node at cluster level
14:                        The node is invalid (attacker node);
15:                        For All of the node at cluster level
16:                            Should be done Authentication Phase;
17:                        End for
18:                End for
19:        Else
20:            Node is Accepted
21:        End while
22: Step 3. General Authentication in the Network
23:         RSA key generation by Base Station (BS);
24:         Key distribution using by the interlock protocol;
25:         Node authentication is done using by node reply;
26:             If the node is invalid then
27:                   The node is rejected.
28:             Else
29:                   The node is Accepted
30:             End if
31: End Procedure
```

---

## 4.5 | The process of preventing Denial of Sleep attacks in ASDA-RSA

To deploy nodes in the network and authenticate them as valid nodes, we generate public and private keys for communications of each sensor node. This is done using the RSA algorithm for key generation. The reason for the use of the RSA algorithm is that in all symmetric key encryption algorithms, the sender and receiver of the message should know the encryption key. When the message sender uses a unique and secret key for encryption, and message recipients use the same key for decryption, exposing a key from one of the message recipients endangers everyone's security. The RSA algorithm is a public key method. This method is the first trusted method among other encryption methods, and is one of the biggest enhancements to cryptography. In public key algorithms, two completely different keys are used for encryption and decryption: the public key and the private key. The public key is used to encrypt information, and everyone knows it, because this key is used only to encrypt the information and the enemies with it will not be able to decrypt the encrypted data by others. The private key is the key by which the encrypted data is decrypted. This key is not even known to friends and trustees. Thus, any network-level entity (whether user, machine, or process) requires two independent keys, only one of which is sensitive and must be kept secret. The nature of this cryptographic algorithm is such that in practice the private key cannot be deduced from the public key. Public and private keys can be exchanged using the interlocking protocol to protect these communications against attacks.

## 4.6 | Node authentication at cluster level

In our proposed method, two levels of authentication are used. One is performed by the cluster head to examine the node members of the cluster and the other generally by the base station. At the first level, we consider a threshold value for $SYN$ packets. During normal operation, if the $SYN$ is lower than the threshold, $SYN$ is used without authentication. If the threshold is exceeded, there is a possibility of a Denial of Sleep attack, and so the authentication must be used with $SYN$ authentication token. The $SYN$ package is authenticated to synchronize the active time interval when the S-MAC algorithm is working. The $SYN$ packet must carry its own identifier in the package as the first field and before the sleep time. The head cluster node checks with the receipt of the $SYN$ packet whether the packet sender is located in its own cluster. If the sender does not exist in this cluster, the $SYN$ packet will not be accepted. This is the first level of authentication. If the initial level of authentication is successful, the head cluster node that is received the $SYN$ packet checks the node's $SYN$ arrival time interval. To do this, each head cluster node keeps the $SYN$ arrival time of the cluster members in its memory. If this interval is exceeded from the threshold, then the suspicion of a Denial of Sleep attack is expected. Here we cannot block the suspicious node because the attack may also have been carried out by an outsider who knows the node behavior and wants to disrupt our entire network operation by sending a packet with a different ID. In our proposed solution we go to an authentication mode. This mode is activated by the head cluster node, which has detected suspicious behavior due to the exceeding from the desired threshold. The head cluster node sends a $SYN-A$ packet to all cluster member nodes to take them to the authentication mode and forces them to send the authentication token in the next $SYN$ token. This continues until it receives the NO $SYN-A$ packet. A cluster member node can transmit it to other nodes in its cluster by receiving $SYN-A$ and if found it exceeding from the threshold. Thus, our proposed method will be able to narrow the attack range and make authentication only compulsory for that area. In the area where the authentication mode is activated, the cluster member nodes must send the $SYN$ packet with an authentication token to the head cluster. When the node is in the authentication mode if it feels that the time interval is lower than the desired threshold it can return to the normal mode. In this case, the NO $SYN-A$ packet will be sent to the member nodes of its cluster and will inform them that they can return to the normal mode, which will reduce the computational overhead.

## 4.7 | General authentication in the network

Therefore, the overall process is that after performing clustering and choosing the appropriate head clusters and cluster-level authentication, the base station will store and have access to the information of all sensor nodes, including the head clusters, as well as all the lower nodes which has been received through head clusters. At the next level, for node authentication, the base station acts as the third party for the negotiations. Public and private keys created by the RSA algorithm are exchanged by the interlocking protocol. The sender node of sleep synchronization messages is a proxy node and the receiving node acts as the evaluator. Each node has a unique private key, which we refer to here as P. The proxy node and the evaluator node share the public key. The base station sends the public key of the proxy node from the base station whenever the evaluator node requests. Instead of sending a direct key, the base station calculates a value of F = P2 mod G. Here P is the private key and G is the public key. The value of F is sent whenever the evaluator requests. By receiving a node response, the key decrypts at the receiver base station and, if validated, the node is accepted. In the ASDA-RSA proposed approach, the incorrect authentication by the attacker becomes difficult and thus preventing Denial of Sleep attack by the attacker and changing the message or forging the SYNC message. The flowchart of proposed methodology is given in Fig. 3.

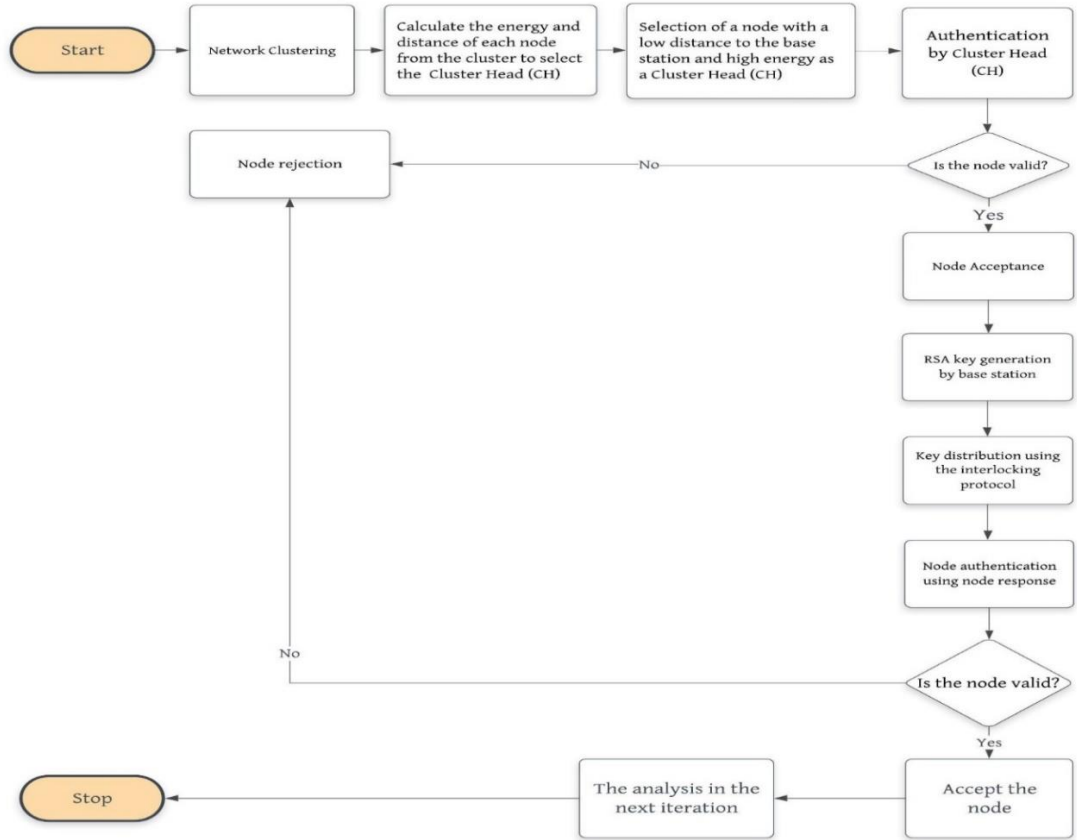

**FIGURE 3** Flowchart of the ASDA-RSA proposed model

## 5 | Evaluating the Performance

In the following section, the performance of our proposed ASDA-RSA approach is evaluated to prevent Denial of Sleep attacks.

### 5.1 |Performance metrics

In this section, the effectiveness and performance of our proposed ASDA-RSA approach is thoroughly evaluated with comprehensive simulations. The results are compared with CrossLayer, ASDA-BlowFish, ASDA-AES, GA-DoSLD, ASDA-3DES, and ASDA-DES approaches proposed in [27, 35, 36] and [29, 37], respectively. The average throughput, Packet Delivery Ratio (PDR), network lifetime, detection ratio and residual energy are evaluated. Notations utilised here are listed in Table 1.

**TABLE 1** Meaning of notations

| Parameters | Description |
|---|---|
| T | Average throughput |
| PDR | Packet delivery rate |
| DR | Detection rate |
| RE | Residual energy |
| NL | Network lifetime |
| $X_i$ | Demonstrate the Number of packets received by sensor node $i$ |
| $Y_i$ | Demonstrate the Number of packets sent by sensor node $i$ |
| $P_S$ | Demonstrate the Size of packet (Kbps). |
| $S_P$ | Denotes the Stop time of simulation |
| $S_T$ | Denotes the Start time of simulation |

1) Average Throughput

Average throughput is the division of the sum of packets sizes received at the destination sensor node, to the difference of simulation stop and start time. Eq. (10) obtains the average throughput for N experiments, and is calculated in Kilobits per second [38-45].

$$T = \frac{1}{n} * \frac{\sum_{i=1}^{n} X_i * P_s}{S_p - S_T} * \frac{8}{1000} \tag{10}$$

2) Packet Delivery Ratio

PDR is the division of the total data packets received at the destination sensor node, to the total number of data packets transmitted by the source sensor node, described in percentage. The average PDR obtained for $N$ experiments is demonstrated by Eq. (11) [31, 46-48].

$$PDR = \frac{1}{n} * \frac{\sum_{i=1}^{n} X_i}{\sum_{i=1}^{n} Y_i} * 100\% \tag{11}$$

3) Network Lifetime

According to the definition, the network lifetime is the elapsed time between of communication and sensing commencement with the receiver, and the time in which the final communication link from active node to the receiver is broken. Network lifetime for all active nodes currently in communication with the receiver is the life time aggregate for all the mentioned nodes at any time instance. If the network is clustered, the network lifetime is the total lifetime for all $CH$ s. Eq. (12) demonstrated the calculation of the $NL$ value.

$$NL = \sum_{i=1}^{m} LCH_i \qquad \text{Where} \qquad LCH_i \text{ is the lifetime of } i\text{th } CH. \tag{12}$$

4) Average Residual Energy

As demonstrated in Eq. (13), unconsumed energy in the node in an arbitrary time instance is the excessive energy maintained in the node following a concluded communication with the receiver. Examples of residual energies are the energy for transmission, energy for reception, wasted energy in the system ($E_{sys}$), fading effects, etc. The parameters exploited for the average residual energy are listed in Table 2.

**TABLE 2** Parameters used for average residual energy

| Parameters | Description |
|---|---|
| $di_0$ | Reference distance larger than the Fraunhofer-distance |
| $di$ | The distance on which the packet is transmitted |
| $Lb$ | demonstrates the number of bits per packet (BPP) |
| $di^2$ | Refers to the power loss of free space channel model |
| $di^4$ | Power loss of multi-path fading channel model |
| $E_{elec}$ | Amount of energy getting dissipated during transmission or reception |
| $lb\hat{I}fsi$ | Transmission efficiency |
| $lb\hat{I}mpi$ | Condition of the channel |

$$Energy_{residual} = Energy_{initial} - \{ET_X + ER_X + E_{sys}\} \quad (13)$$

Where

$$ET_X(lb, di) = \{lbE_{elec} + lb\varepsilon_{fs}di^2, di < di_0\} \quad (14)$$

$$= \{lbE_{elec} + lb\varepsilon_{mpi}di^4, di \geq di_0\}$$

Energy consumed during packet transmission $ET_X(lb, di)$ and packet reception ($ER_X$) are calculated using Eq. (14), and Eq. (15), respectively.

$$ER_X = lbE_{elec}. \quad (15)$$

If $di > di_0$, multipath fading effect occurs, and energy is wasted during transmission. However, since the fading scheme is out of the scope of this paper, the distance is considered to be lesser than the Fraunhofers distance. Moreover, the information for channel state is not considered, while transmission efficiency is considered to be 1.

5) Detection Ratio

As defined, detection ratio is the division of the number of misbehaving DoS nodes detected, to the total number of actual nodes misbehaving. In other words, detection ratio is the probability of successfully identifying all the security threats. Thus, the formulation is shown in Eq. (16). Table 3 lists the parameters employed for detection rate. Therefore, the $FPR, FNR, TPR,$ and $TNR$ is defined as illustrated in Eq. (17), Eq. (18), Eq. (19) and Eq. (20).

**TABLE 3** Parameters used for detection rate

| Parameters | Description |
|---|---|
| True Positive Rate (TPR) | The ratio of normal sensor node that were correctly detected as a normal node. |
| False Positive Rate (FPR) | The ratio of normal sensor node to total normal sensor that were mistakenly detected as a DoS attack. |
| True Negative Rate (TNR) | Correctly rejected. The ratio of malicious sensor node that were correctly detected as a malicious node. |
| False Negative Rate (FNR) | The ratio of malicious sensor node to total normal node that were mistakenly detected as a normal node. |

$$DR = \left(\frac{TPR}{TPR+FNR}\right)*100 \quad (16)$$

$$FPR = \left(\frac{FPR}{FPR+TNR}\right)*100 \quad (17)$$

$$FNR = \left(\frac{TPR+TNR}{All}\right)*100; \quad All = FPR+FNR+TPR+TNR \quad (18)$$

$$TPR = \left(\frac{TPR}{TPR+FNR}\right)*100 \quad (19)$$

$$TNR = \left(\frac{TNR}{TNR+FPR}\right)*100 \quad (20)$$

### 5.2 | Simulation Setup and Comparing Algorithms

The difficulties in implementation and debugging routing protocols in real networks, raises the necessity to consider simulations as a fundamental design tool. The main advantage of simulation is simplifying analysis and protocol verification, mainly in large-scale systems [40, 49-51]. It is possible to employ a NAM in the NS-2 to visualise the results. In this section, the performance of our proposed approach is evaluated using NS-2 as the simulation tool, and the results are discussed further [52-58]. Moreover, the proposed ASDA-RSA are compared with CrossLayer, ASDA-BlowFish, ASDA-AES, GA-DoSLD, ASDA-3DES, and ASDA-DES models. It is worth mentioning that all ASDA-RSA, CrossLayer, ASDA-BlowFish, ASDA-AES, GA-DoSLD, ASDA-3DES, and ASDA-DES parameters and settings are considered to be equal.

### 5.3 | Simulation results

We have simulated ASDA-RSA approached in the NS-2 on Linux Fedora 10 [63-65]. The setting of simulation parameters is given in Table 4.

**TABLE 4** Setting of simulation parameters

| Parameters | Value |
|---|---|
| Sensor network size (m x m) | $80 \times 80\ m^2$ |
| Simulation time | $70\ s$ |
| Number of nodes | 300 |
| Duty cycle | $20-time\ slots$ |
| Transmission range | $150-250\ m$ |
| Packet size | $512\ Bytes$ |
| RTS, CTS, ACK size | $30\ Bytes$ |
| $E_{elec}$ | $100 nJ\ /\ bit$ |
| $\varepsilon_{fsi}$ | $20 pJ\ /\ bit\ /\ m^2$ |
| $\varepsilon_{mpi}$ | $.0015 pJ\ /\ bit\ /\ m^4$ |
| The Initial energy | $35\ J$ |
| The Idle power | $41\ mW$ |
| The Receiving power | $45\ mW$ |
| The Sleep power | $25\ \mu W$ |
| The Transmission power | $41\ mW$ |
| Sensor node sensing power | $5*10^{-8}\ J$ |

Table 5-9 compares the performance of ASDA-RSA with that of CrossLayer, ASDA-BlowFish, ASDA-AES, GA-DoSLD, ASDA-3DES, and ASDA-DES in terms of average throughput, PDR, network lifetime, detection ratio and residual energy.

**TABLE 5** DR of seven approaches with varying degree of malicious sensor nodes

| Misbehaving sensor ratio | Detection rate (%) | | | | | | |
|---|---|---|---|---|---|---|---|
| | ASDA-RSA | CrossLayer | ASDA-Blowfish | ASDA-AES | GA-DoSLD | ASDA-3DES | ASDA-DES |
| 0 | 100 | 100 | 100 | 100 | 100 | 100 | 100 |
| 0.05 | 98 | 95 | 93 | 87 | 85 | 81 | 78 |
| 0.15 | 97 | 91 | 88 | 85 | 80 | 77 | 72 |
| 0.25 | 96 | 85 | 80 | 77 | 75 | 70 | 65 |
| 0.35 | 95 | 81 | 78 | 75 | 72 | 68 | 61 |
| 0.45 | 94 | 75 | 73 | 70 | 65 | 60 | 54 |
| 0.55 | 93 | 71 | 69 | 63 | 60 | 55 | 51 |
| 0.65 | 91 | 68 | 65 | 58 | 52 | 47 | 42 |
| 0.75 | 88 | 63 | 60 | 48 | 45 | 40 | 35 |

**TABLE 6** T of seven approaches with varying degree of malicious sensor nodes

| Misbehaving sensor ratio | Average throughput (in kbps) | | | | | | |
|---|---|---|---|---|---|---|---|
| | ASDA-RSA | CrossLayer | ASDA-Blowfish | ASDA-AES | GA-DoSLD | ASDA-3DES | ASDA-DES |
| 0 | 900.85 | 900.85 | 900.85 | 900.85 | 900.85 | 900.85 | 900.85 |
| 0.05 | 870.5 | 750.4 | 732 | 721 | 700.8 | 630 | 540 |
| 0.15 | 835.4 | 610.9 | 600 | 585 | 580.7 | 502 | 490 |
| 0.25 | 800.2 | 580.4 | 571 | 560 | 550.7 | 390 | 324 |
| 0.35 | 780.9 | 490.3 | 462 | 448 | 420.03 | 301 | 287 |
| 0.45 | 773.6 | 440 | 410 | 404 | 400 | 285 | 245 |
| 0.55 | 760 | 412 | 400 | 370 | 350 | 247 | 218 |
| 0.65 | 703 | 401 | 386 | 362 | 310 | 234 | 208 |
| 0.75 | 620 | 380 | 345 | 327 | 302 | 218 | 196 |

**TABLE 7** PDR of seven approaches with varying degree of malicious sensor nodes

| Misbehaving sensor ratio | PDR (in %) | | | | | | |
|---|---|---|---|---|---|---|---|
| | ASDA-RSA | CrossLayer | ASDA-Blowfish | ASDA-AES | GA-DoSLD | ASDA-3DES | ASDA-DES |
| 0 | 100 | 100 | 100 | 100 | 100 | 100 | 100 |
| 0.05 | 94 | 92 | 90 | 85 | 82 | 80 | 78 |
| 0.15 | 91 | 85 | 82 | 79 | 71 | 78 | 73 |
| 0.25 | 89 | 81 | 78 | 71 | 65 | 75 | 68 |
| 0.35 | 88 | 78 | 75 | 62 | 60 | 66 | 58 |
| 0.45 | 85 | 70 | 66 | 58 | 55 | 52 | 50 |
| 0.55 | 81 | 62 | 58 | 52 | 51 | 46 | 41 |
| 0.65 | 75 | 58 | 51 | 48 | 43 | 40 | 34 |
| 0.75 | 71 | 51 | 47 | 44 | 38 | 32 | 28 |

**TABLE 8** $Energy_{residual}$ of seven approaches with varying degree of malicious sensor node

| Misbehaving sensor ratio | Residual energy (in %) | | | | | | |
|---|---|---|---|---|---|---|---|
| | ASDA-RSA | CrossLayer | ASDA-Blowfish | ASDA-AES | GA-DoSLD | ASDA-3DES | ASDA-DES |
| 0 | 100 | 100 | 100 | 100 | 100 | 100 | 100 |
| 0.05 | 95.7 | 81.7 | 79 | 76 | 75.8 | 72 | 70 |
| 0.15 | 91.4 | 78.4 | 76 | 72 | 71.8 | 69 | 68 |
| 0.25 | 87.4 | 74.2 | 73 | 69 | 67.8 | 68 | 67 |
| 0.35 | 82.8 | 71.08 | 69 | 67 | 65.8 | 64 | 61 |
| 0.45 | 80.7 | 65.1 | 65 | 64.8 | 64.1 | 60 | 58 |
| 0.55 | 77.4 | 60.9 | 60.1 | 59.4 | 58.7 | 54 | 50 |
| 0.65 | 76.1 | 55.3 | 54 | 53 | 51.7 | 49 | 45 |
| 0.75 | 74.3 | 51.7 | 49 | 47 | 43.9 | 41 | 40 |

**TABLE 9** *NL* of seven approaches with varying degree of malicious sensor node

| Misbehaving sensor ratio | Lifetime (in sec) | | | | | | |
|---|---|---|---|---|---|---|---|
| | ASDA-RSA | CrossLayer | ASDA-Blowfish | ASDA-AES | GA-DoSLD | ASDA-3DES | ASDA-DES |
| 0 | 1000 | 1000 | 1000 | 1000 | 1000 | 1000 | 1000 |
| 0.05 | 954 | 878 | 801 | 700 | 690 | 630 | 520 |
| 0.15 | 904 | 790 | 770 | 680 | 610 | 500 | 480 |
| 0.25 | 897 | 710 | 701 | 670 | 500 | 470 | 430 |
| 0.35 | 887 | 670 | 660 | 650 | 450 | 440 | 400 |
| 0.45 | 860 | 610 | 600 | 590 | 401 | 360 | 345 |
| 0.55 | 851 | 590 | 585 | 580 | 364 | 339 | 310 |
| 0.65 | 834 | 570 | 565 | 560 | 312 | 258 | 240 |
| 0.75 | 812 | 560 | 555 | 550 | 304 | 230 | 210 |

Table 10 represents average values of various frameworks for all metrics under Denial of Sleep attack.

**TABLE 10** Average values of various frameworks for all metrics under Denial of Sleep attack

| Schemes | | Throughput | PDR | Detection rate | Residual energy | Lifetime |
|---|---|---|---|---|---|---|
| ASDA-RSA | Number of sensor nodes | 813.681111 | 96.1555556 | 92.8388889 | 94.2444444 | 834.555556 |
| | Misbehaving sensor ratio. | 782.7166667 | 86 | 94.66666667 | 85.08888889 | 888.7777778 |
| | Simulation times | 910 | 91.55555556 | 91.22222222 | 92.22222222 | 46.99 |
| | Attack interval | 659 | 90 | 94.7666667 | 87.6755556 | 868.333333 |
| CrossLayer | Number of sensor nodes | 697.1044444 | 91.0666667 | 86.6488889 | 86.53 | 566 |
| | Misbehaving sensor ratio. | 551.7611111 | 75.22222222 | 81 | 70.93111111 | 708.6666667 |
| | Simulation times | 597.6666667 | 86.88888889 | 82.11111111 | 82.07777778 | 37.9 |
| | Attack interval | 387.54 | 83.44444444 | 77.1111111 | 77.2088889 | 652.333333 |
| ASDA-Blowfish | Number of sensor nodes | 664.8888889 | 89.11111111 | 85.33333333 | 84.87777778 | 550.2222222 |
| | Misbehaving sensor ratio. | 534.0944444 | 71.88888889 | 78.44444444 | 69.45555556 | 693 |
| | Simulation times | 517.3333333 | 85 | 77.44444444 | 78.87777778 | 34.63333333 |
| | Attack interval | 335.4444444 | 75.22222222 | 75 | 74.64444444 | 639.5555556 |
| ASDA-AES | Number of sensor nodes | 653 | 81 | 84.25555556 | 83.64444444 | 539.7777778 |
| | Misbehaving sensor ratio. | 519.7611111 | 66.55555556 | 73.66666667 | 67.57777778 | 664.4444444 |
| | Simulation times | 493.1111111 | 80.35555556 | 75 | 77.44444444 | 32.88888889 |
| | Attack interval | 312.5555556 | 72.66666667 | 73.94444444 | 73.63333333 | 634.4444444 |
| GA-DoSLD | Number of sensor nodes | 638.12 | 78.43333333 | 82.55 | 82.24444444 | 524.5555556 |
| | Misbehaving sensor ratio. | 501.6755556 | 62.77777778 | 70.44444444 | 66.62222222 | 514.5555556 |
| | Simulation times | 459.6666667 | 79.11111111 | 73 | 75.78666667 | 31.38888889 |
| | Attack interval | 300.7133333 | 67.66666667 | 71.44444444 | 72.89888889 | 627 |
| ASDA-3DES | Number of sensor nodes | 368.8888889 | 76.22222222 | 80.06666667 | 79.84444444 | 481.6666667 |
| | Misbehaving sensor ratio. | 411.9833333 | 63.22222222 | 66.44444444 | 64.11111111 | 469.6666667 |
| | Simulation times | 382.7777778 | 76.07777778 | 67.88888889 | 71.44444444s | 24.35555556 |
| | Attack interval | 261.1111111 | 63.88888889 | 68.04444444 | 66.26666667 | 612.4444444 |
| ASDA-DES | Number of sensor nodes | 349.1111111 | 72.66666667 | 78.12222222 | 77.4 | 428.333333 |
| | Misbehaving sensor ratio. | 378.7611111 | 58.88888889 | 62 | 62.11111111 | 437.2222222 |
| | Simulation times | 329.2222222 | 73.73333333 | 60.44444444 | 68.77777778 | 19.63333333 |
| | Attack interval | 221.1111111 | 60.88888889 | 65.21111111 | 62.68888889 | 601.8888889 |

Figure 4 and 5, demonstrates the comparison between our proposed ASDA-RSA scheme, CrossLayer, ASDA-BlowFish, ASDA-AES, GA-DoSLD, ASDA-3DES, and ASDA-DES models in terms of throughput in A, Number of sensor nodes. B, Misbehaving sensor ratio. C, Simulation times. and D, Attack interval. All frameworks are analysed against Denial of Sleep attacks in terms of performance. Since throughput is a vital parameter in sensor networks, we evaluated this parameter to evaluate our method. Since in the ASDA-RSA method the length of the encrypted key is high, it has caused the key not be easily broken by the attacker. Therefore, a larger number of packets reach the destination with unmanipulated information and no packet is eliminated. This very reason has increased the throughput of our method. In the ASDA-Blowfish, ASDA-AES, ASDA-3DES, and ASDA-DES methods, the low length of the key has made infiltrating the key easy and therefore decreased throughput. Plus, in Figure 4 and 5 reveals that our method performs better with different parameters. To provide some explanation on these results, it is worth mentioning that in our proposed method, the node that is selected for the cluster head is the node best in major criteria such as the average energy remaining and its distance to the receiver. Moreover, since the transmission power is set in our proposed method, the packets are guaranteed to arrive at the destination; hence, the number of the packets lost will be reduced, and the proposed method has a higher throughput compared to the current approaches. As shown in Figure 4, and 5, ASDA-RSA increases the throughput by more than 20%, 24%, 26%, 29%, 35% and 43% those of CrossLayer, ASDA-BlowFish, ASDA-AES, GA-DoSLD, ASDA-3DES, and ASDA-DES models, respectively.

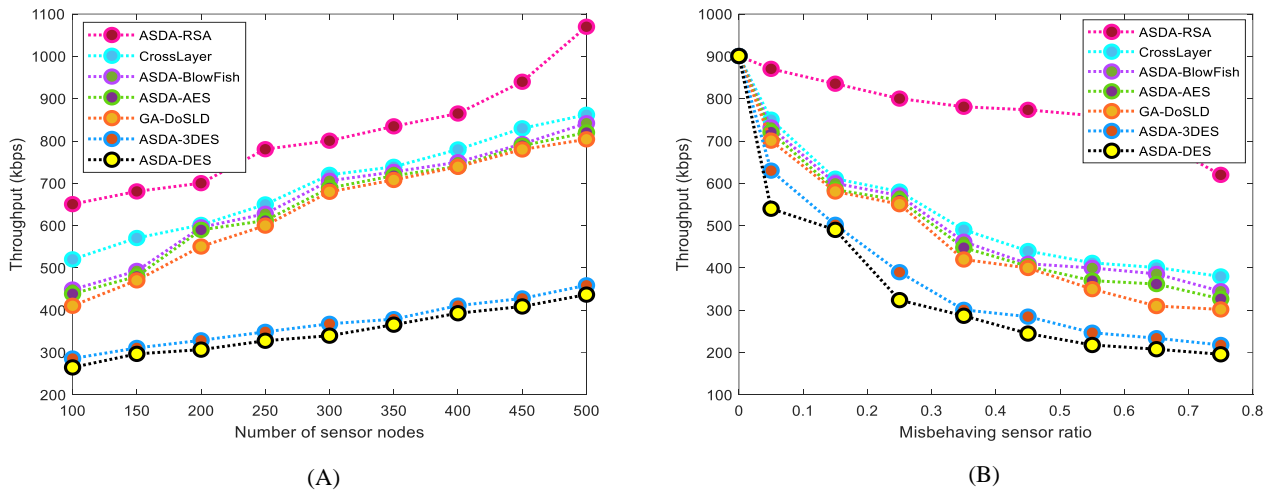

**FIGURE 4** Comparison of the ASDA-RSA proposed scheme, CrossLayer, ASDA-BlowFish, ASDA-AES, GA-DoSLD, ASDA-3DES, and ASDA-DES models in term of Throughput: A, Number of sensor nodes; B, Misbehaving sensor ratio

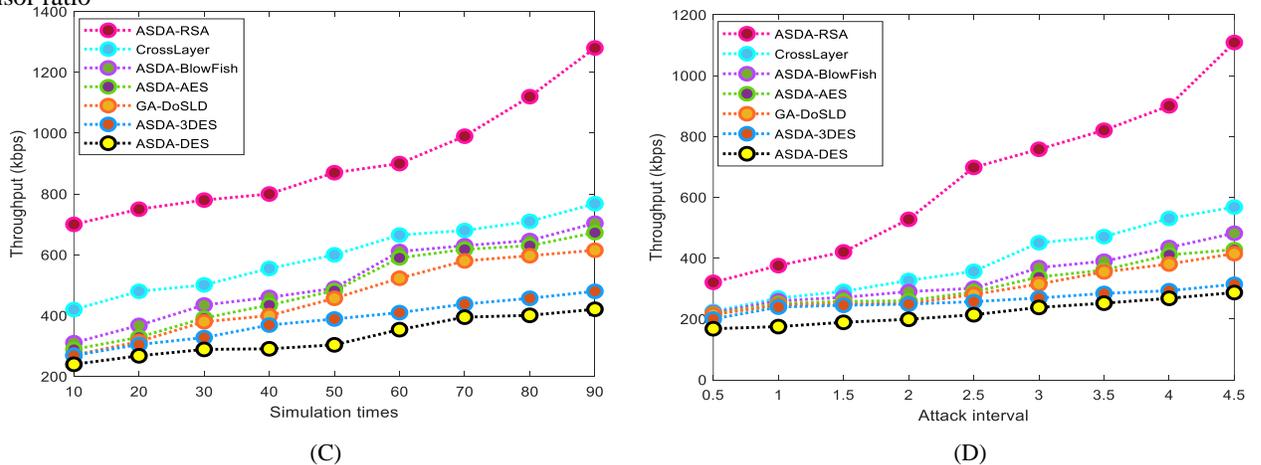

**FIGURE 5** Comparison of the ASDA-RSA proposed scheme, CrossLayer, ASDA-BlowFish, ASDA-AES, GA-DoSLD, ASDA-3DES, and ASDA-DES models in term of Throughput: C, Simulation times; D, Attack interval.

Figure 6 and 7, compares our proposed ASDA-RSA scheme with CrossLayer, ASDA-BlowFish, ASDA-AES, GA-DoSLD, ASDA-3DES, and ASDA-DES models in term of PDR in A, Number of sensor nodes. B, Misbehaving sensor ratio. C, Simulation times. and D, Attack interval. One major issue raised in WSNs, which we focused on in this paper, is the node residual energy, which affects the network lifetime directly. As depicted in Fig. 6, the PDR in ASDA-RSA is higher compared to current approaches. Since the low-key length has caused the key in ASDA-Blowfish, ASDA-AES, ASDA-3DES, and ASDA-DES methods to be easily unlocked by attackers, the packets will be manipulated or will not reach the destination safely. Therefore, this will lead to a decrease in the PDR rate. As shown in Figure 6, and 7, ASDA-RSA increases the PDR by more than 8%, 11%, 13%, 18%, 22% and 26% those of CrossLayer, ASDA-BlowFish, ASDA-AES, GA-DoSLD, ASDA-3DES, and ASDA-DES models, respectively.

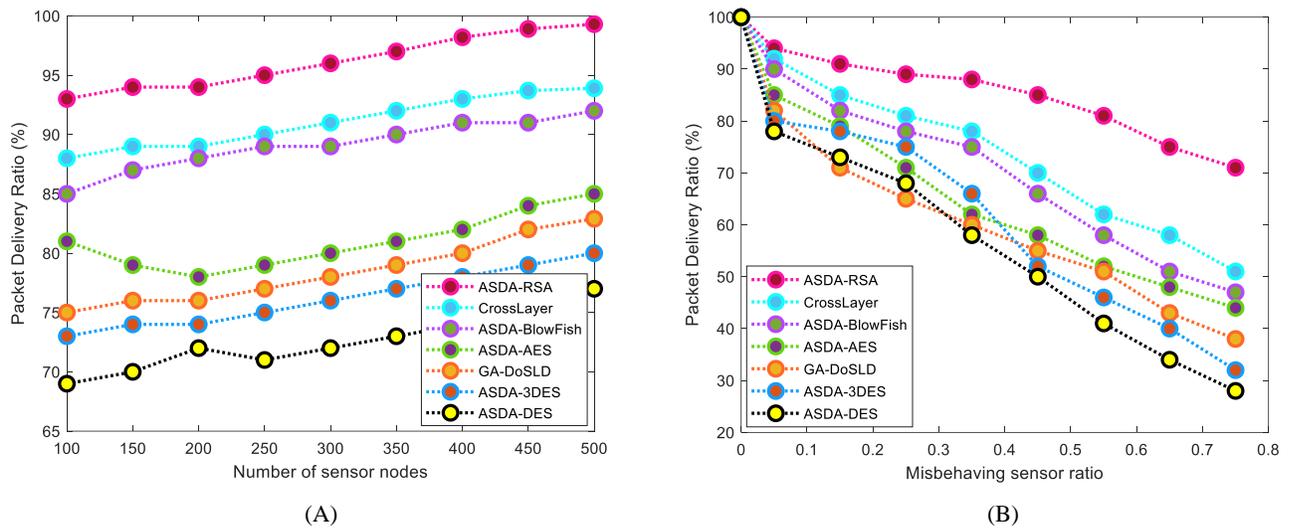

**FIGURE 6** Comparison of the ASDA-RSA proposed scheme, CrossLayer, ASDA-BlowFish, ASDA-AES, GA-DoSLD, ASDA-3DES, and ASDA-DES models in term of PDR: A, Number of sensor nodes; B, Misbehaving sensor ratio.

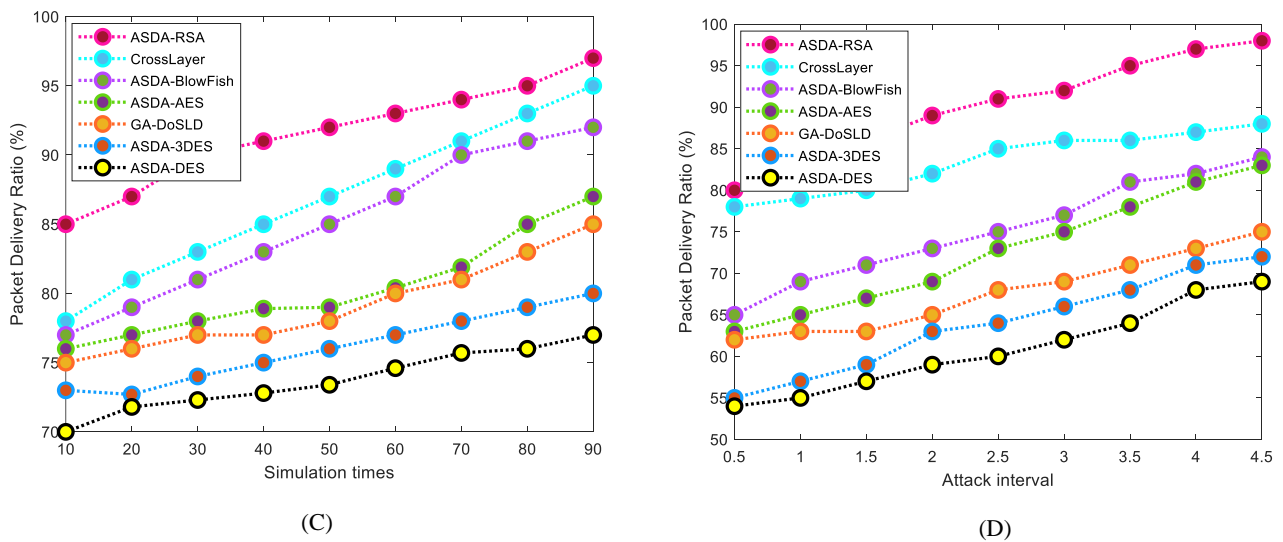

**FIGURE 7** Comparison of the ASDA-RSA proposed scheme, CrossLayer, ASDA-BlowFish, ASDA-AES, GA-DoSLD, ASDA-3DES, and ASDA-DES models in term of PDR: C, Simulation times; D, Attack interval.

In Figure 8 and 9, comparison of our proposed ASDA-RSA framework is illustrated against current approaches. Figure 8A shows that for the detection rate, our proposed ASDA-RSA approach possesses superiority over the contemporary schemes. The main reason is that the proposed algorithm detects denial of sleep attack and performs isolation from the network over the attacks, thereby increasing the detection ratio occurring due to the denial of sleep. For instance, with the GA-DoSLD approach, the detection rate during denial of sleep for 100 to 200 sensor nodes is roughly 82%, with the ratio increasing proportional to the number of the sensor nodes. In addition, by introducing the proposed algorithm to the network with the same sensor nodes, the detection ratio is increased to approximately 93 percent, compared to 87 percent for CrossLayer. The attack detection rate in ASDA-Blowfish, ASDA-AES, ASDA-3DES, and ASDA-DES methods is lower due to their short key length and easily breakable key; because once a packet is hacked, the attacker does not give the network the chance to detect the attack anymore and the packet gets eliminated. In the ASDA-RSA method, because of the long key length, the infiltration rate of the perpetrator is lower and the key does not break easily and therefore, the detection rate is higher. As can be seen, ASDA-RSA algorithm increased detection ratio by more than 11, 14, 17, 19, 23, and 27 percent against CrossLayer, ASDA-BlowFish, ASDA-AES, GA-DoSLD, ASDA-3DES, and ASDA-DES models, respectively.

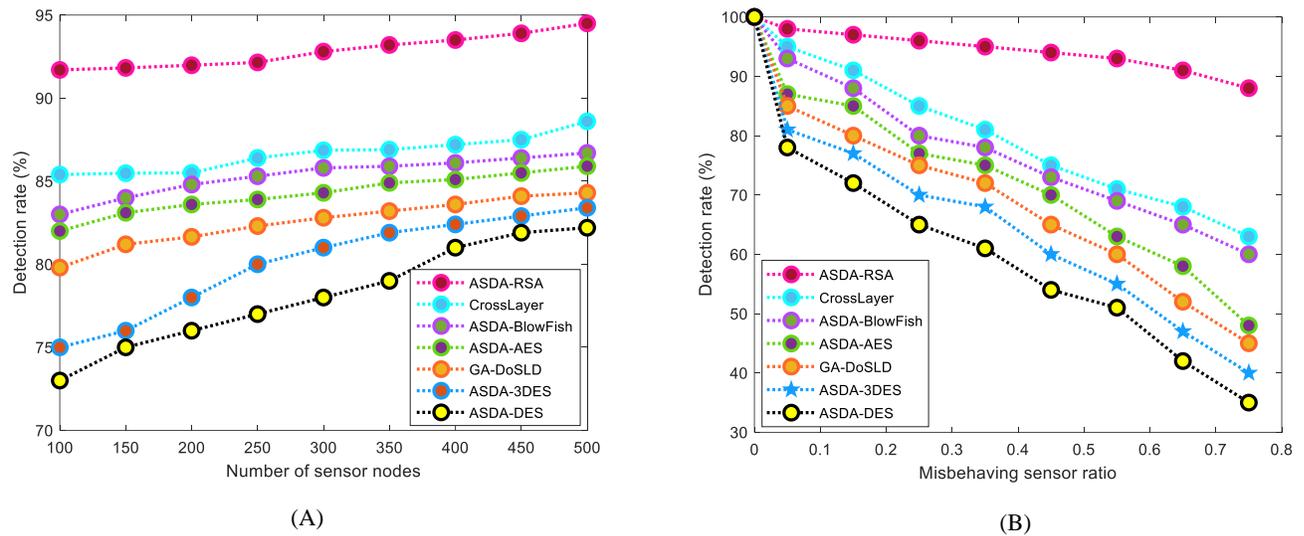

**FIGURE 8** Comparison of the ASDA-RSA proposed scheme, CrossLayer, ASDA-BlowFish, ASDA-AES, GA-DoSLD, ASDA-3DES, and ASDA-DES models in term of Detection rate: A, Number of sensor nodes; B, Misbehaving sensor ratio

Figure 9C illustrates the detection rate against simulation time in the malicious nodes. What is implied here by simulation time is that the packets in the network are randomly transmitted among nodes at arbitrary times, i.e., each node transmits at a different time. Investigating the simulation time is crucial to understanding the ASDA-RSA performance in a Denial of Sleep attack. As indicated in the results, using traditional technique, the detection ratio while under Denial of Sleep attack at time 30 is around 80 percent, decreasing to roughly 68 percent at time 70 against simulation time. However, applying the proposed ASDA-RSA algorithm to the network, while evaluating simulation time, improves the network performance by 10 percent than that of the existing techniques at time 30, and 20 percent at time 70. Finally, in Figure 9D the Denial of Sleep attack detection ratio with attack intervals is demonstrated, aiming to investigate the ASDA-RSA flexibility in the WSN network. As can be seen, using traditional technique, the DoS attack detection rate with 1.5 sec intervals is approximately 65 percent, which increases to approximately 80 percent as the attack interval increases to 4 seconds.

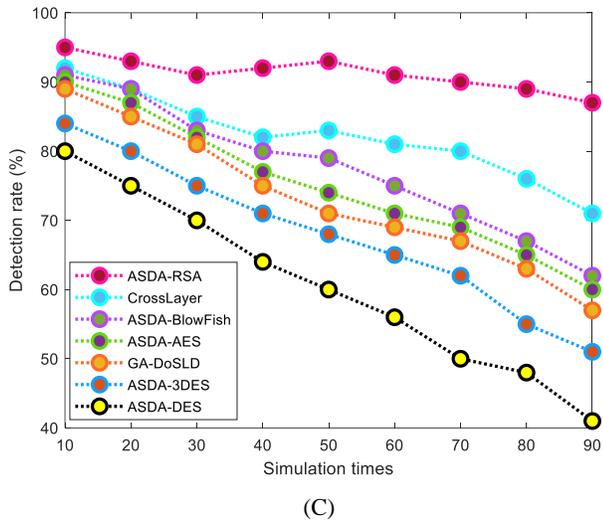
(C)

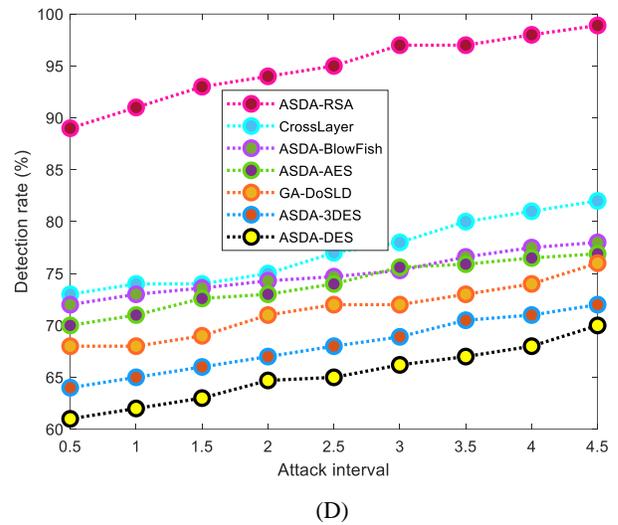
(D)

**FIGURE 9** Comparison of the ASDA-RSA proposed scheme, CrossLayer, ASDA-BlowFish, ASDA-AES, GA-DoSLD, ASDA-3DES, and ASDA-DES models in term of Detection rate: C, Simulation times; D, Attack interval.

Figure 10 and 11 compares the performance of ASDA-RSA with that of CrossLayer, ASDA-BlowFish, ASDA-AES, GA-DoSLD, ASDA-3DES, and ASDA-DES for residual energy. A, Number of sensor nodes. B, Misbehaving sensor ratio. C, Simulation times. And D, Attack interval, respectively. As shown in the figure, ASDA-RSA increases the residual energy by more than 11, 14, 15, 16, 20, and 23% those of current approaches, respectively.

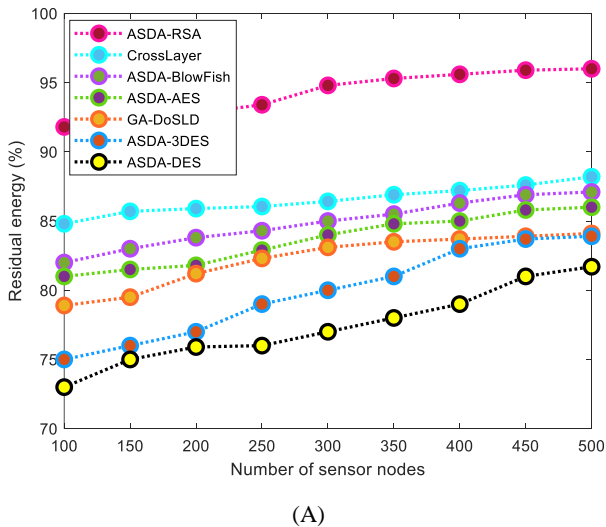
(A)

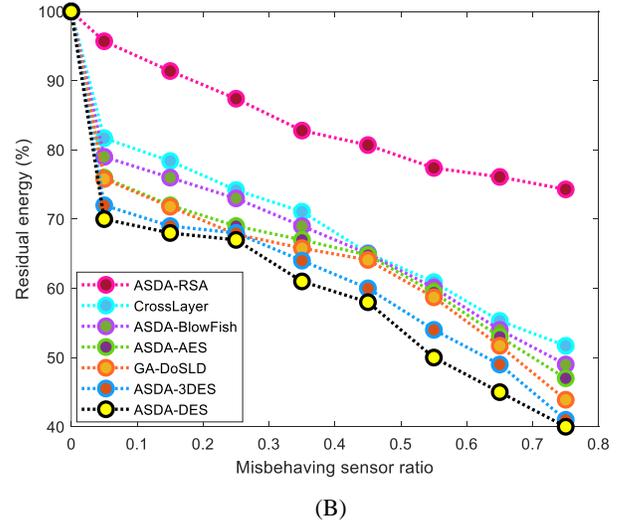
(B)

**FIGURE 10** Comparison of the ASDA-RSA proposed scheme, CrossLayer, ASDA-BlowFish, ASDA-AES, GA-DoSLD, ASDA-3DES, and ASDA-DES models in term of Residual energy: A, Number of sensor nodes; B, Misbehaving sensor ratio

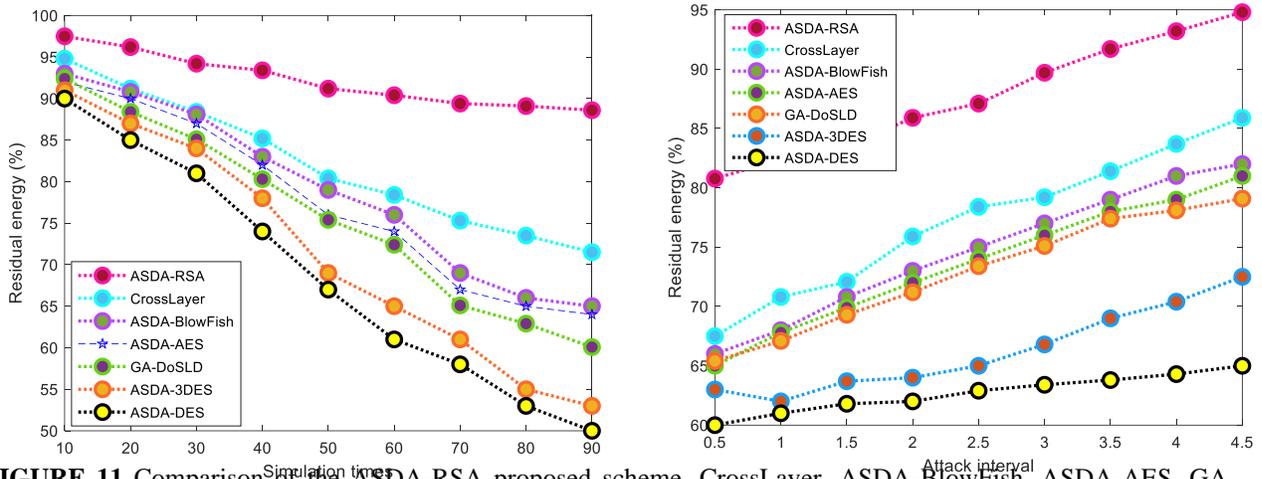

**FIGURE 11** Comparison of the ASDA-RSA proposed scheme, CrossLayer, ASDA-BlowFish, ASDA-AES, GA-DoSLD, ASDA-3DES, and ASDA-DES models in term of Residual energy: C, Simulation times; D, Attack interval.

Figure 12 and 13, shows the comparison of the ASDA-RSA proposed scheme, CrossLayer, ASDA-BlowFish, ASDA-AES, GA-DoSLD, ASDA-3DES, and ASDA-DES models in term of Lifetime. A, Number of sensor nodes. B, Misbehaving sensor ratio. C, Simulation times. And D, Attack interval. Respectively. As mentioned above, the amount of energy consumed in the network and sensor nodes directly affects the lifetime of the WSNs. In this study, we considered the network lifetime as the energy depletion of all nodes of the network. Initially, all nodes are alive and the network has a lifetime of 1000. Over time and energy consumption in the network and in sensor nodes, the total energy of the nodes in the network is terminated and the nodes die, thereby reducing the lifetime of the network. As you can see in Fig. 12 and 13, the proposed ASDA-RSA is better than the curent approaches in terms of network lifetime. In the proposed schema, the sensor node can set the best transmission power. In addition, by selecting the clusters head with the most remaining energy, less overhead and the smallest distance to the sink, the lifetime of the network has been increased. As shown in the Figure 12(a), and 13 ASDA-RSA increases the lifetime by more than 19, 22, 24, 28, 31, and 39% those of current approaches, respectively.

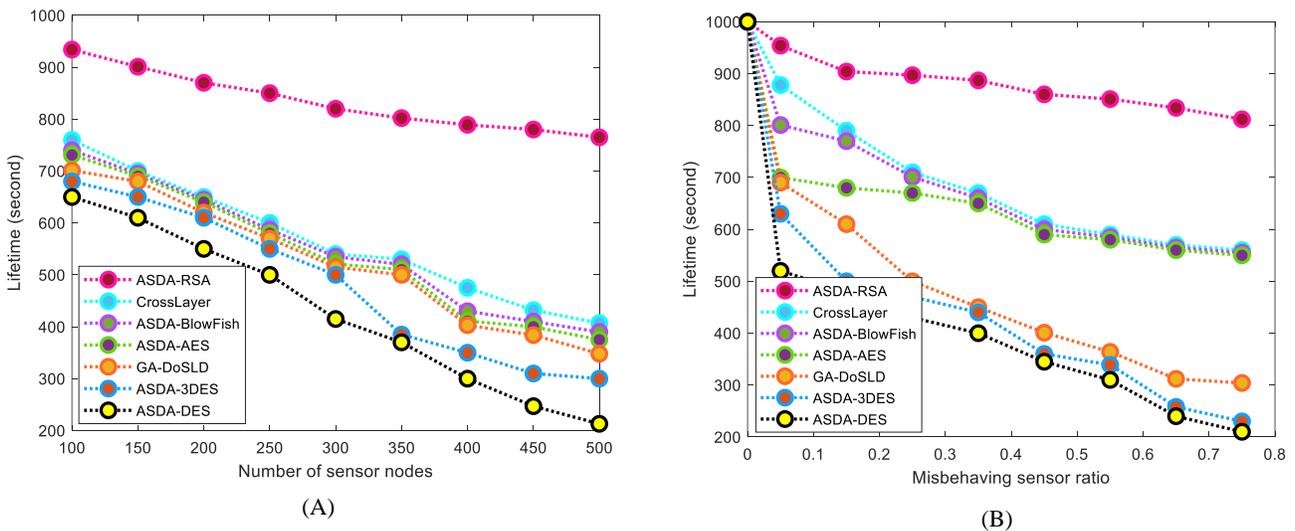

**FIGURE 12** Comparison of the ASDA-RSA proposed scheme, CrossLayer, ASDA-BlowFish, ASDA-AES, GA-DoSLD, ASDA-3DES, and ASDA-DES models in term of Lifetime: A, Number of sensor nodes; B, Misbehaving sensor ratio

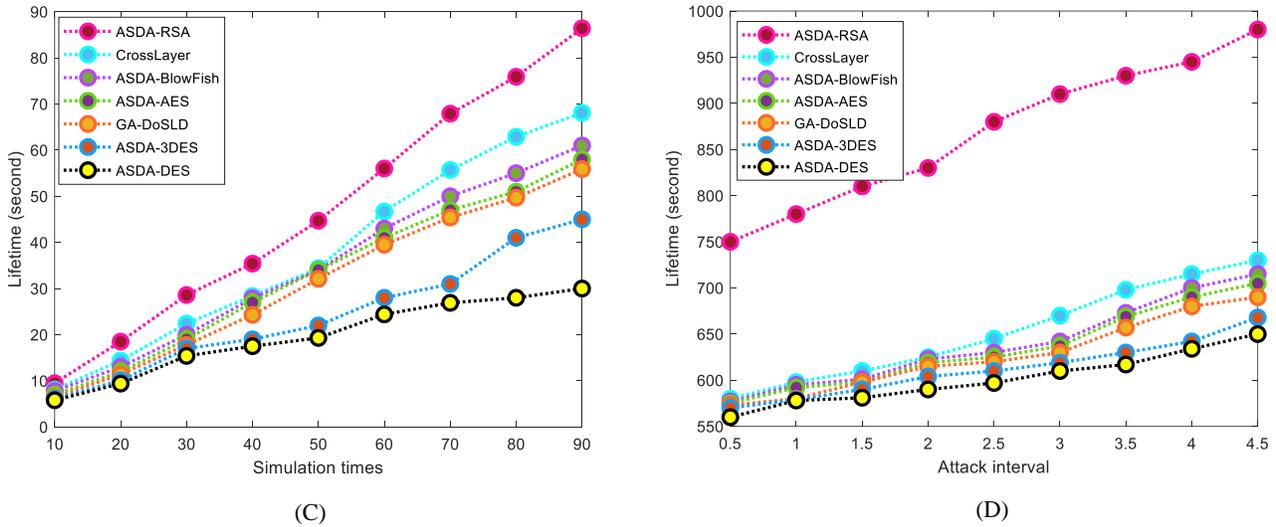

**FIGURE 13** Comparison of the ASDA-RSA proposed scheme, CrossLayer, ASDA-BlowFish, ASDA-AES, GA-DoSLD, ASDA-3DES, and ASDA-DES models in term of Lifetime: C, Simulation times; D, Attack interval.

## 6 | CONCLUSION AND FUTURE WORK

Various properties of wireless sensor networks like limited power source, density, low bandwith, memory with small size, and limited energy make this type of network vulnerable to DoS attacks. Also, limited battery life has turned energy consumption into one of the main challenges in these networks. In this paper, the proposed ASDA-RSA schema to protect the network against DoS attack is discussed in detail. To reduce the energy consumed and improve the network lifetime, in ASDA-RSA a clustering approach based on energy and distance is used to select the proper cluster head. Plus, the RSA cryptography algorithm and an interlocking protocol are employed along with an authentication method to prevent DoS attacks. Simulation results demonstrated that the ASDA-RSA was very robust against DoS attacks. it presented that it has a high level of security and high average throughput (more than 29.5%), PDR (more than 16.33%), network lifetime (more than 27.16%), detection rate (more than 18.5%), and average residual energy (more than 16.5%) as compared to current approaches.

In this paper, we presented a method for reducing energy consumption by countering denial-of-sleep attacks based on cryptography and authentication. In the proposed method, all sinks were fixed and this (network with fixed sinks) makes the nodes adjacent to the sink act as the cluster head or an intermediary for the transmission of other nodes' data to the sink. In turn, this leads to the rapid reduction of the energy of these nodes and therefore decreased network lifetime. to improve this matter, in future works we intend to use mobile sinks based on one of the meta-heuristic algorithms which is a technique for improving energy consumption and increasing network lifetime.

### ORCID

*Reza Fotohi*  http://orcid.org/0000-0002-1848-0220

## REFERENCES

1. Juneja V, Gupta DV. Security Against Vampire Attack in ADHOC Wireless Sensor Network: Detection and Prevention Techniques. InInternational Conference on Wireless Intelligent and Distributed Environment for Communication 2018 Aug 18 (pp. 25-38). Springer, Cham.